\documentclass[useAMS,usenatbib]{mn2e}
\usepackage{natbib}
\usepackage{amsmath,amsfonts}
\usepackage{epsfig}
\usepackage{lscape}
\usepackage{mathptmx}

\def\reff@jnl#1{{\rm#1\/}}

\def\aj{\reff@jnl{AJ}}                  
\def\araa{\reff@jnl{ARA\&A}}            
\def\apj{\reff@jnl{ApJ}}                        
\def\apjl{\reff@jnl{ApJ}}               
\def\apjs{\reff@jnl{ApJS}}              
\def\ao{\reff@jnl{Appl.Optics}}         
\def\apss{\reff@jnl{Ap\&SS}}            
\def\aap{\reff@jnl{A\&A}}               
\def\aapr{\reff@jnl{A\&A~Rev.}}         
\def\aaps{\reff@jnl{A\&AS}}             
\def\azh{\reff@jnl{AZh}}                        
\def\baas{\reff@jnl{BAAS}}              
\def\jrasc{\reff@jnl{JRASC}}            
\def\memras{\reff@jnl{MmRAS}}           
\def\mnras{\reff@jnl{MNRAS}}            
\def\pra{\reff@jnl{Phys. Rev. A}}         
\def\prb{\reff@jnl{Phys. Rev. B}}         
\def\prc{\reff@jnl{Phys. Rev. C}}         
\def\prd{\reff@jnl{Phys. Rev. D}}         
\def\prl{\reff@jnl{Phys. Rev. Lett}}      
\def\pasp{\reff@jnl{PASP}}              
\def\pasj{\reff@jnl{PASJ}}              
\def\qjras{\reff@jnl{QJRAS}}            
\def\skytel{\reff@jnl{S\&T}}            
\def\solphys{\reff@jnl{Solar~Phys.}}    
\def\sovast{\reff@jnl{Soviet~Ast.}}     
\def\ssr{\reff@jnl{Space~Sci.Rev.}}     
\def\zap{\reff@jnl{ZAp}}                        
\def\nat{\reff@jnl{Nature}}             

\def\p#1by#2{{\partial{#1} \over \partial{#2}}}
\def\pp#1by#2#3{{\partial^2{#1} \over \partial{#2}\partial{#3}}}
\def\d#1by#2{{{\rm d}{#1} \over {\rm d}{#2}}}
\def\dd#1by#2#3{{{\rm d}^2{#1} \over {\rm d}{#2}{\rm d}{#3}}}

\title[Constraints on spinning dust towards Galactic targets with the VSA]
{Constraints on spinning dust towards Galactic targets with the VSA:
a tentative detection of excess microwave emission towards 3C396}

\label{firstpage}
\author[A. Scaife et~al.] 
{Anna Scaife$^1$,
 David A. Green$^1$,
 Richard A. Battye$^2$,
 Rod D. Davies$^2$, 
\newauthor 
 Richard J. Davis$^2$, 
 Clive Dickinson$^{3,6}$, 
 Thomas Franzen$^1$,
 Ricardo G{\'e}nova-Santos$^1$, 
\newauthor 
 Keith Grainge$^1$, 
 Yaser A. Hafez$^2$, 
 Michael P. Hobson$^1$, 
 Anthony Lasenby$^1$, 
 \newauthor 
 Guy G. Pooley$^1$,
 Nutan Rajguru$^{4}$,
 Rafael Rebolo$^{5}$,
 Jos\'e Alberto Rubi\~no-Martin$^{5}$, 
\newauthor
 Richard D. E. Saunders$^1$, 
 Paul F. Scott$^1$, 
 David Titterington$^1$,  
 Elizabeth Waldram$^1$,
\newauthor
 Robert A. Watson$^2$. 
 \vspace{0.03in}\\
$^1$  Astrophysics Group, Cavendish Laboratory, 19 J. J. Thomson Ave, Cambridge CB3 0HE\\
$^2$ Jodrell Bank Observatory, Macclesfield, Cheshire SK11 9DL \\
$^3$ Jet Propulsion Laboratory, 4800 Oak Grove Drive, M/S 169-327,
Pasadena, CA 91109, USA.\\
$^4$ University College London, Department of Physics \& Astronomy, Astrophysics Group, Gower Street, London WC1E 6BT\\
$^5$ Instituto de Astrof{\'i}sica de Canarias, 38200 La Laguna, Tenerife, Spain.\\ 
$^6$ California Institute of Technology, Mail Code 105-24, Pasadena, CA 91125, USA.\\
}

\date{Accepted ---; received ---; in original form \today}
\pagerange{\pageref{firstpage}--\pageref{lastpage}}
\pubyear{2006}

\begin{document}
\maketitle

\begin{abstract}
We present results from observations made at 33 GHz with the Very
Small Array (VSA)
telescope towards potential candidates in the Galactic plane for
spinning dust emission. In
the cases of the diffuse {\sc Hii} regions LPH96 and NRAO591 we find no
evidence for anomalous emission and, in combination with Effelsberg
data at 1.4 and 2.7 GHz, confirm that their spectra are consistent with optically
thin free--free emission. In the case of the infra-red bright SNR 3C396 we find
emission inconsistent with a purely non-thermal spectrum and discuss
the possibility of this excess arising from either a spinning dust
component or a shallow spectrum PWN, although we conclude that the
second case is unlikely given the strong constraints available from
lower frequency radio images.
\end{abstract}


\section{Introduction}

A localized excess of emission in the microwave region was first detected in
the $\it{COBE}$/DMR data and was initially attributed to free--free
emission (Kogut et al. 1996a, 1996b). Since then this
anomalous emission has been detected by a number of authors (de
Oliveira-Costa et al. 2002, 2004; Banday et al. 2003; Finkbeiner et
al. 2004; Watson et al. 2005; Fern{\'a}ndez-Cerezo et~al. 2006), and has been nicknamed
`$\it{Foreground}$  $\it{X}$'. Although initially ascribed to thermal
bremsstrahlung in view of its strong correlation with thermal dust, low
H$\alpha$ surface brightness measurements (Leitch et al. 1997) implied
gas temperatures in excess of $10^6$\,K, which were rejected on energetic grounds by Draine \&
Lazarian (1998a). Its physical mechanism has yet to be
constrained; the most popular interpretation is that of rapidly
rotating dust grains, or $\it{spinning}$ $\it{ dust}$ (Draine \& Lazarian 1998a, 1998b). Other
mechanisms which have been proposed include magnetic dust emission
(Draine \& Lazarian 1999), flat spectrum synchrotron (Bennett et
al. 2003b), and bremsstrahlung from very hot electrons (Leitch et al. 1997).
 Spinning dust causes an excess of emission in the 10 -- 50\,GHz region of the
spectrum, where the combination of synchrotron, bremsstrahlung and
thermal dust emission is a minimum, and is problematic especially
for CMB experiments which utilise this region to minimise
foreground contamination and avoid atmospheric emission. Consequently the presence of a poorly constrained foreground
such as this anomalous dust emission is a potentially serious problem for CMB
observers and needs to be better understood in order to be 
correctly removed. This has led to several directed observations
(Finkbeiner et~al. 2004; Watson et~al. 2005; Casassus et~al. 2004, 2006; Dickinson et~al. 2006)
being made towards targets suggested by the theoretical predictions of
Draine \& Lazarian (1998b; hereinafter DL98b). Whilst these have been mainly directed at
{\sc Hii} regions and dark clouds, DL98b also suggest that it may be
possible to detect $10 - 100$\,GHz emission from spinning dust in
photodissociation regions. Here we present observations at 33\,GHz
towards two {\sc Hii} regions and one supernova remnant (SNR) for
which infrared observations imply the presence of significant
photodissociation regions (PDRs). 

\section{The Telescope}\label{telescopes}

The VSA is a 14 element interferometer sited at the Teide Observatory, in Tenerife, at an altitude
of 2400\,m. The VSA operates in a single 1.5\,GHz wide channel at a central
frequency of 33\,GHz (Scott et~al. 2003). The 14 antennas use HEMT
amplifiers with typical system temperature $\approx$35\,K. In its new
super-extended configuration the VSA uses mirrors of diameter 65\,cm.  The horn-reflector antennas
are mounted on a tilt table hinged east--west and each antenna
individually tracks the observed field by rotating its horn axis
perpendicularly to the table hinge,  wavefront coherence being
maintained with an electronic path compensator system (Watson et~al. 2003).  

The individual tracking of the VSA antennas allows for the filtering
of contaminating signals. These may be celestial sources, such as the
Sun and Moon, or ground-spill and other environment based spurious signals. The VSA also uses a ground-shield to
minimise ground-spill. The consequences of this design 
are two-fold. First, the VSA is able to observe continuously and
can filter out emission from the Sun
and Moon when they are as close as $9^{\circ}$.  Second, the VSA is
unaffected by ground-spill contamination for fields within
35$^{\circ}$ of the zenith and so is able to make direct images of the
sky, rather than employing the lead--trail approach of many
other interferometers operating in the microwave band.

\section{Observations}
\label{sec:lphcomp}
Observations were carried out as single pointings with a primary beam
of 72\,arcmin FWHM.
The data were calibrated using Tau\,A and
Cas\,A in accordance with 
VSA reduction procedures. This calibration, along with appropriate flagging and filtering
of the data, was performed using the special purpose package {\sc reduce}
developed specifically for reduction of VSA
observations. Details of the VSA reduction and calibration procedure
can be found in Dickinson et al (2004) and references therein.

Many surveys at lower frequency do not have similar resolution to the
super-extended VSA, making comparison difficult. Instead we extrapolate from the Effelsberg 100\,m telescope at 2.7\,GHz (F$\ddot{\rm{u}}$rst
et al. 1990). These data\footnote[1]{The Effelsberg survey data were downloaded from the MPIfR sampler
survey website: {\tt http//www.mpifr-bonn.mpg.de/survey.html}} are single dish maps with a circular beam of 4.3\,arcmin FWHM. 

For a robust comparison the 2.7\,GHz Effelsberg maps were
multiplied by the primary beam of the VSA and visibility sampling was
performed in the $\it{uv}$--plane with a correction for the Effelsberg
beam.

\section{LPH96}

The {\sc Hii} region LPH96 (RA = 06$^{\rm{h}}$ 36$^{\rm{m}}$ 40$^{\rm{s}}$, $\delta$ = +10$^{\circ}$ 46$\arcmin$ 28$\arcsec$, J2000) has
been observed with the Green Bank 43m telescope between 5 and 10\,GHz
(Finkbeiner et al., 2002) and was shown to have a rising
spectrum consistent with that expected from spinning dust. However, a
pointed observation made with the CBI telescope (Dickinson et al.,
2006) at 31 GHz shows emission consistent with an approximately flat
spectrum source with only little possibility of spinning dust. 

At 33\,GHz the peak flux density of LPH96 is 1.100$\pm$0.030\,Jy~beam$^{-1}$. The
synthesized beam of the VSA towards LPH96 is
9.1$\times$6.3\,arcmin$^2$ and 
the source is slightly resolved with structure extending towards the
north and southwest. The 2.7\,GHz data after sampling gives a peak flux density of
1.436$\pm$0.010 Jy beam$^{-1}$. Comparing this with the VSA flux density we find a spectral index
$\alpha$ =  0.106$\pm$0.026, where the errors are calculated from the
thermal noise outside the beam on the 2.7 and 33 GHz maps. The
spectral index, $\alpha$, is here defined so that flux density scales as
$\nu^{-\alpha}$. This
calculation fails to take into account systematic errors; taking
errors of 5 percent on the flux density scales at each frequency gives $\alpha$ =  0.106$\pm$0.065.

The VSA measurement shows no indication of the excess emission
which would be expected at 33\,GHz from the warm neutral medium (WNM)
spinning dust model of DL98b 
consistent with the Green Bank data. The result is more consistent
with that found by the CBI telescope 
(Dickinson et al. 2006) who found emission at 31\,GHz consistent only with
$\alpha = 0.06 \pm 0.03$. Dickinson et al. also note that the Galactic
plane survey of Langston et al. (2000) failed to detect LPH96 at 14.35
GHz with a detection limit of 2.5\,Jy.  
\section{3C396}

The supernova remnant (SNR) 3C396 ($=$G39.2$-$0.3), is a shell-like remnant
at
radio frequencies, with a mean angular diameter of $7\farcm8$ (Patnaik et
al.\
1990). Its spectral behaviour has been extensively studied in the radio,
most
notably by  Patnaik, and has a  non-thermal radio spectrum with $\alpha
\approx
0.42$ between $\sim 400$~MHz and $\sim 10$~GHz. Below 30~MHz
catalogued flux densities for 3C396 are contaminated by the near by steep-spectrum pulsar,
PSR
1900$+$0.5 (Manchester \& Taylor 1981). A pulsar wind nebula near the centre
of
this SNR has been detected in X-rays (see Olbert et al.\ 2003), but the
remnant
has not been detected optically. Only a lower limit of  7.7~kpc for its
distance is available from its {\sc Hi} absorption observation (Caswell
et~al.\
1975). Patnaik et~al.\ conclude that the neighbouring {\sc Hii} region
NRAO~591
is likely to be at a distance of $\simeq 14$~kpc.

\subsection{VSA observation of 3C396}
\begin{figure}
\centerline{\includegraphics[height=7.cm,width=7.cm,angle=0]{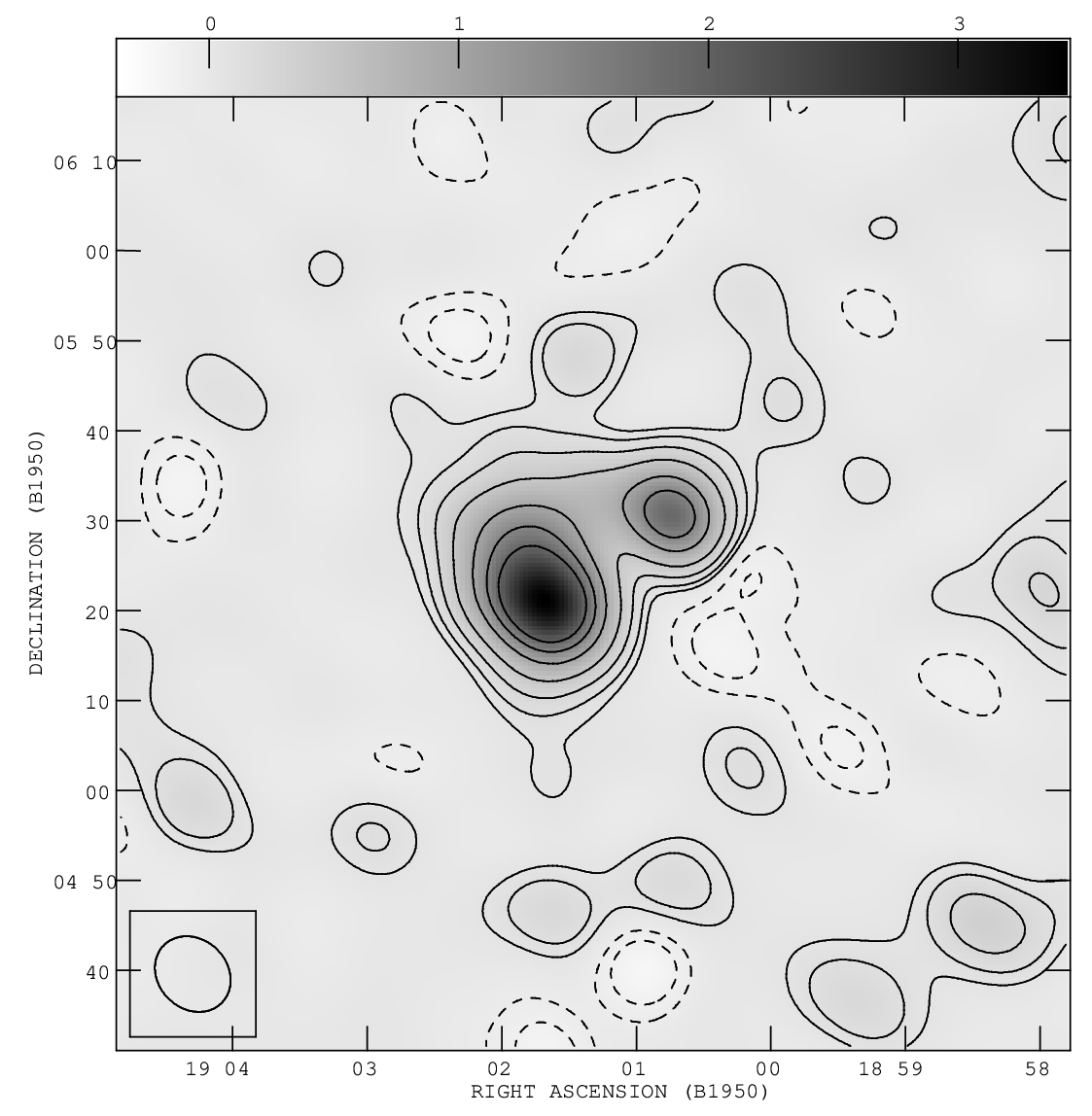}}
\caption{VSA observation of 3C396 with the {\sc Hii} region NRAO 591
also visible to the north--west. Contours are overlaid at $-$2,
$-$1, 1, 2, 4, 8, 16, 32 and 64 $\sigma$.\label{fig:3c396vsa}}
\end{figure}

3C396 was observed in August 2006 in a short observation of 1.8
hours. Mapping and clean-based deconvolution were
performed using the {\small AIPS} package. The observation was not limited by thermal noise but rather
by the dynamic range of the telescope and has rms noise of
34.0\,mJy. The {\small VSA} map is shown in Fig.~\ref{fig:3c396vsa};
it shows both 3C396 and also the {\sc Hii} region NRAO 591 to the north--west
of the remnant. The VSA beam towards 3C396 is
$9.1\times7.7$\,arcmin$^2$. Interesting features include a protrusion to
the north and a faint detection of the ``blow-out'' tail to the
north--east, both of which are also present at longer radio wavelengths. 

Fits were obtained using the
{\small AIPS} task {\small JMFIT} by drawing a bounding box around
both sources and fitting for two Gaussians and a base level. The peak
flux density for 3C396 was found to be 3.21$\pm$0.29\,Jy~beam$^{-1}$
and the integrated flux density 6.64$\pm$0.33\,Jy at 33\,GHz, with the peak at
19$^{\rm{h}}$01$^{\rm{m}}$43$\fs2$ +05$^{\circ}$22$'$22$\farcs 4$ (B1950). For the secondary
source NRAO 591 we find a peak flux density of 1.71$\pm$0.09\,Jy~beam$^{-1}$
and an integrated flux density of 2.40$\pm$0.12\,Jy centered at
19$^{\rm{h}}$00$^{\rm{m}}$47$\fs4$ +05$^{\circ}$31$'$06$\farcs 3$
(B1950). The errors here include contributions from the rms noise on the
observation, the statistical error from the Gaussian fits, and a
conservative 5\% error from the flux calibration which dominates the
overall value. A complete
discussion of the VSA flux calibration may be found in Dickinson
et~al. (2004).

\subsection{The integrated spectrum of 3C396}
Again we use observations from the Effelsberg 100\,m telescope to
examine the emission at lower radio frequencies. Using Effelsberg data
at 1.4 (Reich et~al. 1997) and 2.7\,GHz we convolve the data at 2.7\,GHz to match the 1.4\,GHz
resolution of 9.4\,arcmin and find flux densities of 14.9 $\pm$ 1.3 Jy and 11.4 $\pm$
1.3 Jy at 1.4 and 2.7\,GHz, respectively. These values agree with those of Reich et al. (1990) and give a
spectral index of $\alpha$ = 0.46, which is consistent with the mean
of 0.45 found for shell-type SNRs (Green, 2004; 2006). After
sampling the Effelsberg observation at 2.7\,GHz to match the {\it
uv}--coverage of the super-extended VSA towards 3C396 we find
a flux density of 10.90$\pm$0.51\,Jy. Our best fit index implies a
flux density of 3.4\,Jy at 33\,GHz extrapolated from the Effelsberg data.

\begin{figure}
\centerline{\includegraphics[height=8.cm,width=6.cm,angle=-90]{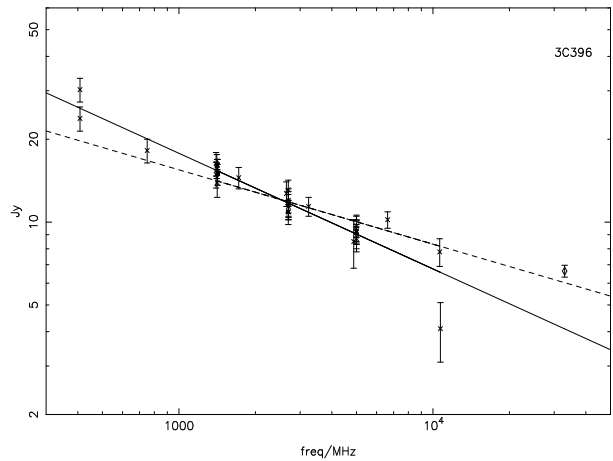}}
\caption{Integrated spectrum of 3C396. Data points are taken from
Patnaik et~al. (1990) with the exception of Effelsberg measurements at
1.4 and 2.7\,GHz which are indicated by filled squares. A spectral
index of $\alpha = 0.42$ is shown as a solid line and of
$\alpha = 0.32$ as a dashed line. The measurement from the VSA at
33\,GHz is shown as an unfilled diamond.\label{fig:3c396spec}}
\end{figure}

We investigate the spectrum of 3C396 using our own flux density and
other published values; all errors are quoted to
1\,$\sigma$. Taken at face value the {\small VSA} 
observation of 3C396 would imply an index $\alpha^{2.7}_{33}$~=~0.20$\pm$0.02, broadly consistent with a region of thermal
emission. Alternatively, if we
assume the index $\alpha^{1.4}_{2.7}$ 
of 0.46 determined from the Effelsberg data, then the {\small VSA}
flux density would imply an excess of emission
seen at microwave frequencies towards this source. A large
number of observations between 400\,MHz and 5\,GHz exist and to
confirm this spectral index we compile a
spectrum using flux densities taken from Patnaik et al. (1990) who made
corrections to the original measurements to bring them onto the flux density scale
of Baars et al. (1977). The integrated spectrum of 3C396 is shown in
Fig.~\ref{fig:3c396spec}. Data from Patnaik et~al. is shown as
crosses, the data from Effelsberg as filled squares and the VSA data
as an unfilled diamond. Performing a weighted least squares fit to the
Patnaik and Effelsberg data we 
find an index of $\alpha = 0.42\pm0.03$. The VSA measurement at
33\,GHz is inconsistent with this spectrum, shown in
Fig.~\ref{fig:3c396spec} as a solid line. Including the VSA
data at 33\,GHz we find a spectral index of $\alpha = 0.32\pm0.02$,
shown as a dashed line. However, $\chi^2$ values for the two fitted spectra
show that the spectral index of 0.42 ($\chi^2_{\rm{red}} = 1.07$, 31
d.o.f., P(1.07) =
0.366) is a better fit to the
data. Indeed a spectral index of 0.32 ($\chi^2_{\rm{red}} = 1.95$, 32
d.o.f., P(1.95)
= 0.001) would be unusually flat for a
supernova remnant with less than 7\,\% of shell like or possible shell like
supernova remnants having a spectral index of less than or equal to
0.32 (Green 2006). However, we hesitate to over-interpret this
statistic since the catalogued spectral indices are by no
means uniform in quality.

\begin{figure}
\centerline{\includegraphics[height=8.cm,width=4.cm,angle=-90]{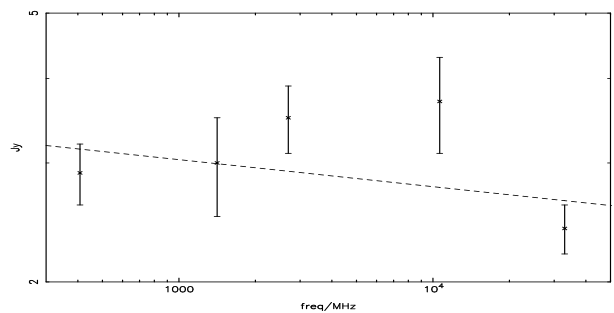}}
\caption{Integrated spectrum of NRAO 591. Fluxes are taken from
Patnaik et~al. (1990) with the exception of the VSA data point at
33\,GHz which is shown as a diamond. A best fit spectral index of
$\alpha = 0.05\pm0.06$ is shown as a dashed line. \label{fig:nraospec}}
\end{figure}

Since previous observations (Finkbeiner et~al. 2004) have suggested the presence of spinning
dust in {\sc Hii} regions we have also compiled a spectrum for NRAO\,591. We use flux densities compiled by Patnaik et al. (1990) which are plotted in
Fig.~\ref{fig:nraospec} and find a spectrum
compatible with that of optically-thin free--free emission.

\begin{figure}
\centerline{\includegraphics[height=7.cm,width=7.cm,angle=0]{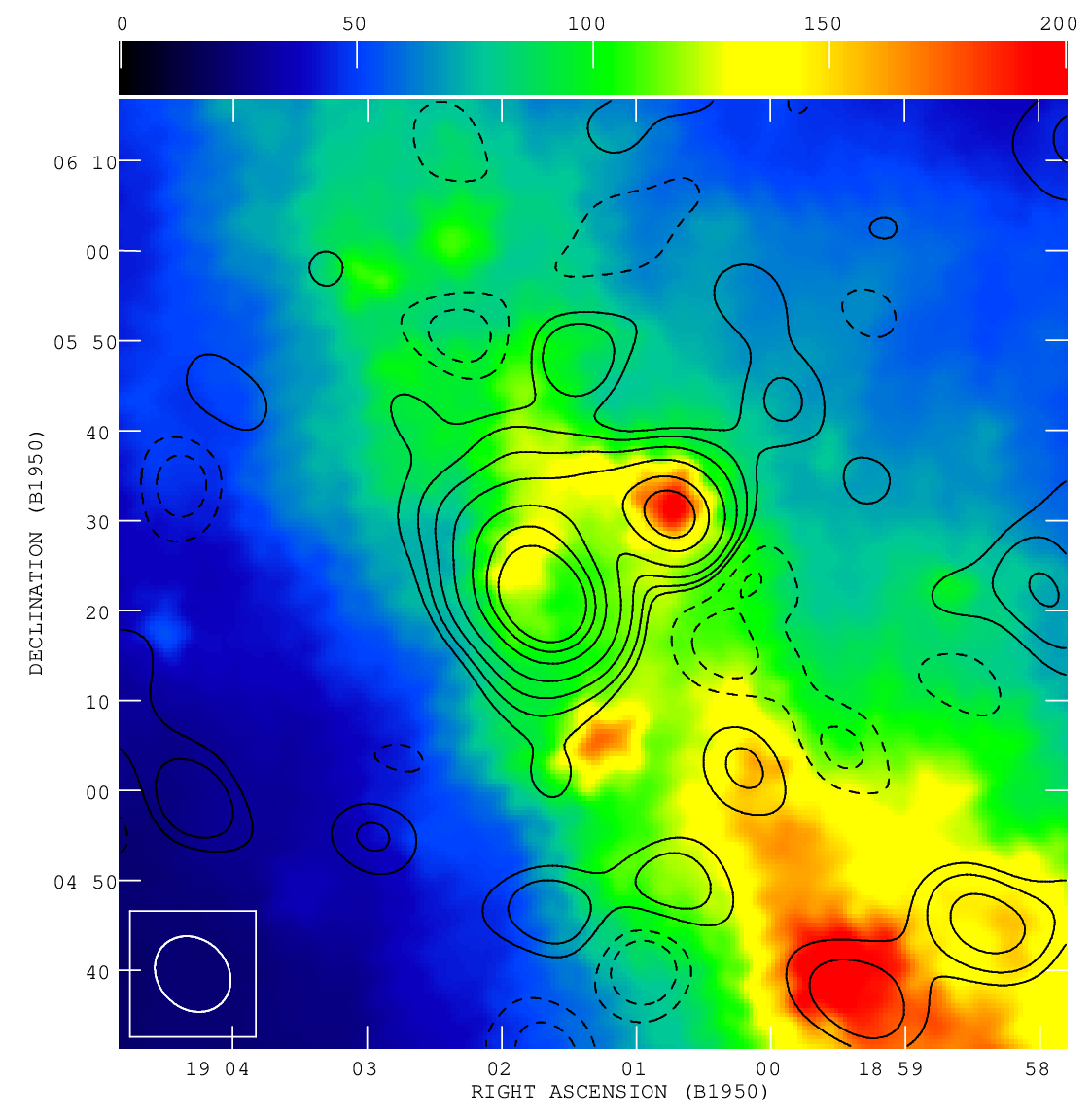}}
\caption{Infra-red emission towards 3C396. Colourscale is IRAS
100\,$\mu$m data overlaid with the VSA observation at
33\,GHz. Contours are $-$2,
$-$1, 1, 2, 4, 8, 16, 32 and 64$\sigma$.\label{fig:100um}}
\end{figure}

\subsection{Discussion and Conclusions}

For the source 3C396 we constrain the possible excess emission relative to the non-thermal
contribution at 33\,GHz by subtracting a non-thermal model
extrapolated from lower frequency. For the flux densities between 400\,MHz
and 10\,GHz, shown in
Fig.~\ref{fig:3c396spec}, a best fit spectral index of $\alpha =
0.42\pm0.03$ was found. This index implies a flux
density of $4.19\pm$0.11\,Jy at 33\,GHz, where the error is
statistical only. This leaves $2.45\pm$0.35\,Jy that
may be due to an anomalous component. If we include a 5 percent error
on the flux calibration these errors are increased to 0.24 and 0.41\,Jy,
respectively. The excess accounts for 37\% of the total 33\,GHz flux
density.  

Radio recombination line observations (Anantharamaiah, 1985) put an upper
limit on the emission measure of 3C396 of 280\,cm$^{-6}$\,pc and fits
an electron temperature of 5000\,K for the gas. This implies an upper limit on
the free--free emission from the SNR of 0.061\,Jy\,beam$^{-1}$ at 33\,GHz. The beam sizes used for this measurement are
poorly matched to that of the VSA and may cause the values to be
over-estimated. Taking this into account it can be seen that the
free--free contribution is small compared to the non-thermal and can
only account for $\sim$2\% of the excess emission. In addition, models
including both free--free and non-thermal contributions provide poor
fits to the Patnaik et~al. data. 

The possibility of the VSA seeing a secondary radio source in projection
towards 3C396 is small. Radio sources with flux densities
$> 100$\,mJy are seen with a frequency of 0.2/deg$^2$. The possibility
of a rising spectrum source at $>$ 5\,GHz reduces this
number by a factor of 10 (Waldram et~al., 2003; Cleary et~al., 2005). 

Olbert et~al. (2003) report the presence of a small pulsar wind nebula 
(PWN) within the 3C396 SNR , although the authors note the lack of any  
corresponding radio feature in high resolution 20\,cm VLA images of
the remnant (Dyer \& Reynolds 1999). For this PWN (or plerion-like
component), to account for the 
excess flux density seen at 33\,GHz it would be very obvious at
1.4\,GHz, where it would have to contribute approximately 1/6 of the
total flux density, assuming a flat spectral index -- Olbert
et~al. however suggest that its contribution is   
$\leq$\,1/25 of the total radio flux density at 1.4\,GHz.

The infra-red emission at 100\,$\mu$m towards 3C396 is shown in
Fig.~\ref{fig:100um}. In this region of the Galactic plane it is not
certain whether the emission is associated with the SNR remnant, or is
merely a projection. Reach et~al. (2006) suggest that higher resolution Spitzer data shows emission at
3.6 to 8\,$\mu$m  which may be associated with the SNR. They find IRAC colours
for these regions consistent with both PDRs and {\sc Hii} regions. 

\begin{figure}
\centerline{\includegraphics[height=8.cm,width=6.cm,angle=-90]{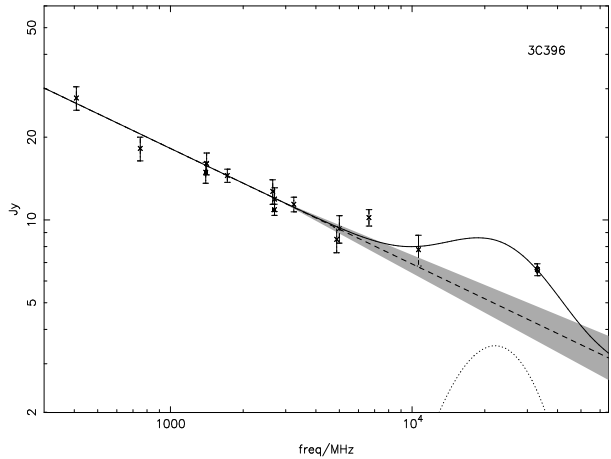}}
\caption{Integrated spectrum of 3C396. Data points from Patnaik and
Effelsberg are as in Figure 3 with data binned at similar frequencies. The best
fit spectral index of $\alpha = 0.42\pm0.03$ is shown as a dashed line and a spinning
dust model (DL98b) scaled to fit the VSA data at
33\,GHz is shown as a dotted line. The measurement from the VSA at
33\,GHz is shown as an unfilled diamond.\label{fig:3c396spec2}}
\end{figure}

If the excess emission seen at 33\,GHz from 3C396 is associated with
these regions then it is possible it may arise from the dipole
emission of Draine \& Lazarian (1998a;1998b). However, the discrepancy
in resolution between the Spitzer and VSA telescopes precludes a more
detailed spatial analysis. This possibility is illustrated in
Figure~\ref{fig:3c396spec2} where the data is shown with the WNM
spinning dust model of DL98b. For clarity we have binned the data of
Patnaik et~al. at similar frequencies, and have excluded one point at
10.7\,GHz. In addition, we note that the WNM model of DL98b,
where the column density towards 3C396 is determined from the full-sky magnitude map of Schlegel, Finkbeiner \& Davis
(1998), predicts a peak 
flux density at 33\,GHz of 3.5\,Jy\,beam$^{-1}$ at the resolution of
the VSA due to spinning dust. This gives an integrated flux density of
approximately 
7\,Jy, using the ratio of peak to integrated flux found with the
VSA. However, it is likely that this method overestimates the
integrated flux density since 
the morphology of the thermal dust emission appears more compact than the
emission seen with the VSA. It is however within a factor of 3 of the
excess emission we see at 33\,GHz.

In conclusion, we have assessed the possibility of spinning dust
emission at 33\,GHz towards the SNR 
3C396. Apart from Cas~A and Tau~A, few SNR have been studied in the
microwave region. Consequently, in order to confirm this 
possibility further measurements are required in the range 10--20\,GHz.

\section*{ACKNOWLEDGEMENTS} 
We thank the staff of the Mullard Radio Astronomy Observatory, the
Teide Observatory and the Jodrell Bank Observatory for their
invaluable assistance in the commissioning and operation of the
VSA. The VSA is supported by PPARC and the IAC. AS acknowledges
the support of a PPARC studentship.

Part of the research described in this paper was carried out at the
Jet Propulsion Laboratory, California Institute of Technology, under a
contract with the National Aeronautics and Space Administration.

We would also like to thank the anonymous referee for his careful
reading of this paper.

\bsp
\label{lastpage}

\end{document}